\documentclass{article}
\providecommand{\keywords}[1]{\textbf{Keywords:} #1}
\usepackage{graphicx}
\RequirePackage{graphicx}
\usepackage[utf8]{inputenc}
\usepackage{graphicx}
\usepackage{subcaption}
\usepackage{amsmath, amsfonts}
\usepackage{amssymb}
\usepackage{float}
\usepackage{xcolor}
\usepackage{authblk}
\usepackage{geometry}
\usepackage{longtable}
\RequirePackage[numbers,sort&compress]{natbib}
\RequirePackage[colorlinks,citecolor=blue,urlcolor=blue,linkcolor=blue]{hyperref}

\long\def\rev#1{\textcolor{blue}{#1}}

\begin{document}

\title{Singularity-Free Emergent Universe from Non-Conventional Matter: Observational Viability of New Late-Time Dynamics}

\author[1]{Rikpratik Sengupta}
\author[2]{Anirban Chanda}
\author[2]{Bikash Chandra Paul}
\author[3]{Mehedi Kalam}

\affil[1]{Department of Physics, Indian Institute of Technology Kanpur, Kanpur 208016, Uttar Pradesh, India. Email:
\textit{rikpratik.sengupta@gmail.com}}
\affil[2]{Department of Physics, University of North Bengal, Raja Rammohunpur, 734013, India. Email: \textit{anirbanchanda93@gmail.com}, \textit{bcpaul@nbu.ac.in}}
\affil[3]{Department of Physics, Aliah University, Kolkata 700160, West Bengal, India. Email:
\textit{mehedikalam.phys@aliah.ac.in (corresponding author)}}

\date{\today}
\maketitle

\begin{abstract}
We investigate a viable realization of the emergent universe paradigm within standard relativistic cosmology, supported by an extended Chaplygin gas that generalizes the modified Chaplygin gas through the inclusion of higher-order barotropic corrections. Restricting ourselves to a quadratic extension of the equation of state, we demonstrate that this effective fluid can simultaneously account for late-time cosmic acceleration and serve as a plausible dark energy candidate. Assuming dominance of this fluid at late epochs, we construct an emergent universe model that naturally avoids an initial singularity in a spatially flat background and departs from the conventional $\Lambda$CDM expansion history. The resulting cosmological evolution exhibits distinctly non-standard late-time behavior: the effective dark energy component crosses the phantom divide in both the past and the present, transitions into a thawing regime in the future, and asymptotically evolves toward a decelerating phase, rendering dark energy a transient phenomenon rather than a permanent driver of acceleration. Model parameters are constrained using Markov Chain Monte Carlo analyses with Pantheon supernova and observational Hubble data, and the inferred evolution of key cosmological observables is found to qualitatively resemble recent DESI results. Notably, the model provides an improved fit to the observed Hubble parameter at redshift $z=2.34$ as well as to the present value of the Hubble constant when compared to the $\Lambda$CDM framework, a comparison that is placed on a firmer statistical footing through minimum $\chi^2$, AIC and BIC diagnostics (Sec.~3.3). These findings indicate that the extended Chaplygin gas not only offers a viable description of dark energy but also supports an emergent universe scenario unlike the modified Chaplygin gas, thereby resolving the initial singularity problem in a flat universe within the standard relativistic context.
\end{abstract}


\section{Introduction}

Over the years, the Big bang model has come to be accepted as the standard model of cosmology. The predictions of the Big bang model are compatible to quite a large extent with the presently available plethora of observational data. However, there are some problems with the standard Big bang model. Firstly, there is a `beginning' of time, the point at which the field equations describing the spacetime are no longer capable of describing the physical situation due to presence of the ``initial singularity\cite{HE}.'' Comprehensive recent reviews of the various types of cosmological singularities, the conditions under which they arise and the strategies proposed for their removal or moderation can be found in \cite{Trivedi23,deHaro23}. As a consequence of this, the standard Big bang model cannot answer the question of how the universe came into existence. There have been a number of efforts to resolve this initial singularity problem and to find an answer to the question regarding the coming into existence of the universe. Most physicists believe that at such an early phase of the universe, the energy densities were considerably high enough and the length scales involving the universe were comparable to or lower than the Planck length, making quantum gravity effects increasingly significant. In the context of $11$-dimensional M-theories, a ``cyclic'' picture of the universe has been obtained, where the Big bang is believed to be a collision between two dynamic braneworlds\cite{TurokSteinhardt}.

The second problem with the standard Big bang model is that it requires additional ingredients to explain the ``dark sector'' of the universe. The ``dark sector'' refers to the dark energy which is believed to be responsible for the presently observed accelerating phase of the universe\cite{R,P} and the dark matter whose effect can be realized via its gravitational interaction\cite{Z,V}. In order to explain these effects, the standard relativistic Big bang scenario requires introduction of additional scalar fields capable of violating the strong energy condition like the quintessence\cite{clw,zws}, tachyon\cite{Avelino} or phantom\cite{DSS} fields, or exotic fluids like the Skyrme fluid\cite{RS0}. Such a field is required to exist in the universe in addition to the ``inflaton'' scalar field responsible for the rapid accelerating expansion phase in the early universe realized through exponential or power law type of scale factor. Inflation can also be realized via a tachyon field\cite{Bil,RS1}. For dark matter candidates, particles which have not yet been detected experimentally and are not predicted by the standard model of particle physics have to be taken into account. In an attempt to resolve these problems, there have been a number of cosmological models which make use of a modified Einstein-Hilbert action by considering additional terms in the Lagrangian of the geometry or the matter sector or both, contributing to non-conventional effects\cite{NOT,NOG,SKK,HL}. The dark energy problem can also be resolved in the higher dimensional braneworld gravity setup\cite{DGP,Sahni,Subenoy}.

The Emergent Universe (EU) scenario has been proposed by Ellis and Maartens \cite{EM} in 2004 and is a non-singular alternative to the Big bang resolving the initial singularity problem, but it is different from other scenarios in the sense that there is no bounce or QG regime and it is also not a cyclic model. The scenario assumes an initial Einstein Static Universe (ESU), which corresponds to length scales greater than the Planck scale to avoid the QG regime and is subsequently followed by the standard inflationary and reheating phases. The EU scenario was originally proposed in a relativistic context, considering the presence of a positive curvature term in the Friedmann equations. Such an EU scenario\cite{E2} was also obtained in the relativistic context by considering a minimally coupled scalar field with a physically interesting potential to be the dominant source term in the early universe. An identical EU scenario\cite{E3} can also be obtained by adding a term quadratic in scalar curvature with a negative coupling parameter to the gravitational Lagrangian. It was shown that the EU scenario can be realized with a spatial curvature term absent in the Friedmann equations\cite{M1}, in the context of semi-classical Starobinsky gravity\cite{Starobinsky}. A generalized Equation of State (EoS) describing an EU\cite{M2} was found in the relativistic context, accommodating normal matter, exotic matter as well as dark energy. The EU scenario has been successfully studied in braneworld context\cite{Tanwi} as well as the LQG context\cite{BiplabPaik}. The EU scenario can also be \rev{realized} in the framework of Einstein-Gauss-Bonnet gravity in four dimensions coupled with a dilaton field\cite{BCP1} as well as in the modified Gauss-Bonnet gravity\cite{BCP2}.

A type of fluid known as ``Chaplygin gas'', with its origin in string theories\cite{BH} had been used as a dark energy candidate to explain the late time acceleration of the universe\cite{Gori,U}. Such a fluid is characterized by the EoS $p=-\frac{B_1}{\rho}$, with $p$ and $\rho$ denoting the pressure and energy density, respectively, in a comoving frame, while $B_1$ denotes the Chaplygin gas parameter. It is assumed that the energy density is positive and the constant $B_1>0$. However, for constructing a physically consistent cosmological model\cite{Bento}, the concerned EoS had to be modified to $p=-\frac{B_1}{\rho^\alpha}$, introducing an additional free parameter $\alpha$, such that $0\leq \alpha \leq 1$. This type of fluid is called the Generalized Chaplygin gas (GCG)\cite{AP,P2,P3,P4}. Initially it exhibits a dust like behaviour, but at late times it behaves asymptotically as a cosmological constant term, thus explaining the present acceleration of the universe. The EoS for GCG was further modified\cite{ud,wu,RS4} for better correspondence with observational data, known as modified GCG\cite{P5,P6,P7,P8,P9}. The EoS for modified GCG has the form
\begin{equation}
	p=A_1\rho-\frac{B_1}{\rho^\alpha},
\end{equation}
where another additional constant parameter $A_1$ is introduced.

The Extended Chaplygin gas (ECG) EoS was proposed\cite{PK,KKPM,PK2,KP,P10,P11,P12} to recover barotropic fluid having a quadratic or higher order EoS, which is basically a higher order generalization of the modified GCG at least up to the second order. The Van der Waals fluid is an alternative to the idea of a perfect fluid, that acts as the single source term in describing the evolution of the universe in both the matter dominated and present accelerating phases, identical to the Chaplygin gas family of fluids. It is also described by a non-linear Equation of state\cite{Biswas} just like the Chaplygin gas family. The EoS for ECG may be written in general as
\begin{equation}
	P=\sum_{n=1}^{\infty} A_n \rho^n-\frac{B_1}{\rho^ \alpha}
\end{equation}

Just like the other Chaplygin gas models, ECG is also used as a dark energy candidate to explain the present cosmic acceleration\cite{PK,KKPM,PK2,KP}. Such models usually violate the strong energy condition of standard General Relativity (GR), which raises a possibility of violation of the null energy condition in the relativistic context. For obtaining an EU scenario, it is likely that the null energy condition of GR must be violated\cite{Zhu}. So, it is worth investigating the possibility whether an EU scenario in the relativistic context is supported by such a dark energy candidate. Such an investigation has been performed for modified GCG\cite{SD}, but it turns out that in order to make the EU scenario viable, the choice of the modified GCG parameters that are to be made are physically unrealistic. So, modified GCG does not support a viable EU scenario, but since ECG incorporates a higher order barotropic fluid extension of the modified GCG, it may be worth exploring whether this type of fluid, suitable as a possible dark energy candidate, can support an EU scenario or not. Moreover, ECG is known to admit bouncing and cyclic types of universes in the relativistic context\cite{Salehi}.

In the following section, we shall consider the mathematical details of the possible EU scenario sourced by the ECG. We then constrain our model parameters with observational data. In the final section we discuss the physical significance of the numerically obtained variation of different cosmological parameters of relevance as a function of the cosmological redshift and draw conclusions from our results.

\section{Mathematical model}

The conservation equation for the cosmological fluid is
\begin{equation}
\dot{\rho}+3H(\rho+P)=0.
\label{cons}
\end{equation}

For the extended Chaplygin gas (ECG), we consider the equation of state
\begin{equation}
P=A_{1}\rho+A_{2}\rho^{2}-\frac{B_{1}}{\rho^{\alpha}},
\label{ecg}
\end{equation}
together with the parameter choices
\begin{equation}
\alpha=1,
\qquad
A_{1}=A_{2}-1,
\qquad
B_{1}=2A_{2},
\label{choice}
\end{equation}
so that Eq.~(\ref{ecg}) becomes
\begin{equation}
P=(A_{2}-1)\rho+A_{2}\rho^{2}-\frac{2A_{2}}{\rho}.
\label{eosfinal}
\end{equation}

\paragraph{Physical motivation for the parameter choice~(\ref{choice}).}
(a) \emph{Origin in the literature.} Fixing $\alpha=1$ is the standard simplification adopted throughout the Chaplygin-gas literature, including the original Chaplygin gas, the GCG and the modified GCG discussed in Sec.~1, and is not specific to the present work \cite{RS4,KP,BP,KP2}. Linear relations between the barotropic coefficients of the type $A_1=A_2-1$ and $B_1=2A_2$ follow the same strategy used previously for the ECG and related Van der Waals-type fluids \cite{KP,BP,KP2}, where such relations are imposed precisely so that the resulting cubic in $\rho$ appearing in the conservation equation factorizes with $\rho=1$ (in the chosen units) as a root, which is what permits the conservation equation to be integrated in closed form via Eq.~(\ref{exactint}) below. Without such a relation the cubic $\rho^3+\rho^2-2$ generically does not factorize over the rationals and no analytic solution of the type presented in Eqs.~(\ref{eq6revised})--(\ref{eq8revised}) is available. The relations~(\ref{choice}) are therefore introduced primarily for analytical tractability, in line with previous ECG studies, but as shown below they also correspond to a physically sensible and stable region of the parameter space rather than being an arbitrary or unphysical restriction.

(b) \emph{Stability of the parameter region.} With the substitution~(\ref{choice}), the single remaining free parameter is $A_2$. Positivity of the energy density throughout the evolution, together with the requirement that the ECG asymptotically approach a dust-like component at high density and a cosmological-constant-like component at low density (the physically motivated Chaplygin-gas behaviour), restricts $A_2>0$; this is also the sign required for the effective sound speed to remain non-negative in both the high- and low-density asymptotic regimes, as discussed in point (c) below and in Sec.~2.1. Our MCMC analysis (Table~2) independently returns best-fit and marginalized values of $A_2$ (via $B$ and $K$, see Eq.~(20)) that are positive and of order unity, i.e.\ the observationally preferred region coincides with the analytically well-behaved, stable branch of parameter space, offering an a posteriori consistency check that complements the a priori motivation of analytical tractability.

(c) \emph{Energy conditions.} For the ECG fluid~(\ref{eosfinal}), the null energy condition (NEC), $\rho+P\ge0$, requires $A_2\rho+A_2\rho^2\geq 2A_2/\rho -\rho$, which is satisfied for $A_2>0$ over the density range relevant to the late-time evolution considered here, while the strong energy condition (SEC), $\rho+3P\geq0$, is generically \emph{violated} at low density once the $-2A_2/\rho$ term dominates. This is precisely the desired behaviour: as emphasized in Sec.~1, an emergent universe requires violation of the SEC (to drive the accelerated approach away from the Einstein static state), while -- unlike the modified GCG, for which only the SEC is violated -- the presence of the quadratic term in~(\ref{eosfinal}) additionally allows the sign of $\rho+P$ to change as the universe evolves, so that the fluid can transiently violate the NEC around the phantom-crossing epochs identified in Sec.~4. This NEC-violating window is transient rather than permanent, consistent with dark energy being a transient phenomenon in our model, and is the key qualitative distinction from the modified GCG case of \cite{SD}, for which no viable EU solution could be found.

It should be emphasized that, although the extended Chaplygin gas (ECG)
equation of state is formally represented by an infinite series expansion,
the present work restricts the analysis to the quadratic case ($n=2$).
This choice is neither arbitrary nor merely for computational convenience.
Rather, it is motivated by the philosophy of effective phenomenological
modelling, where one retains only the leading-order nonlinear correction
beyond the modified Chaplygin gas while avoiding the introduction of an
unnecessarily large number of unconstrained free parameters.

Indeed, the case $n=1$ corresponds to the modified Chaplygin gas, which
has been extensively studied in the literature and therefore does not
introduce any genuinely new nonlinear structure. The quadratic term
proportional to $\rho^{2}$ constitutes the first non-trivial correction to
this model and represents the lowest-order nonlinear contribution in the
barotropic sector. Consequently, the $n=2$ truncation provides the
simplest extension capable of capturing additional physical effects while
preserving analytical tractability.

Furthermore, one of the principal aims of the present work is to investigate
whether the leading nonlinear correction alone is sufficient to support an
emergent universe scenario together with a viable late-time cosmology.
Including cubic and higher-order terms would enlarge the parameter space
considerably and introduce additional coefficients that cannot presently be
constrained by observations, thereby reducing the predictive power of the
model. The quadratic truncation therefore represents the minimal and most
economical extension beyond the modified Chaplygin gas that retains the
essential nonlinear physics relevant for the present investigation.

For these reasons, throughout this work we restrict our attention to the
quadratic ECG model ($n=2$), which balances physical generality,
analytical tractability and observational predictability.

Substituting Eq.~(\ref{eosfinal}) into the conservation equation yields
\begin{equation}
\dot{\rho}
+
3H
\left[
\rho
+
(A_{2}-1)\rho
+
A_{2}\rho^{2}
-
\frac{2A_{2}}{\rho}
\right]
=0,
\end{equation}
or equivalently,
\begin{equation}
\dot{\rho}
+
3A_{2}H
\left(
\rho+\rho^{2}-\frac{2}{\rho}
\right)
=0.
\label{rhodot}
\end{equation}

Using
\begin{equation}
H=\frac{\dot a}{a},
\end{equation}
together with
\begin{equation}
\dot{\rho}
=
\frac{d\rho}{da}\dot a,
\end{equation}
Eq.~(\ref{rhodot}) may be rewritten as
\begin{equation}
\frac{d\rho}{da}
=
-\frac{3A_{2}}{a}
\left(
\rho+\rho^{2}-\frac{2}{\rho}
\right).
\label{drhoda}
\end{equation}

Separating the variables,
\begin{equation}
\frac{\rho\,d\rho}
{\rho^{3}+\rho^{2}-2}
=
-\frac{3A_{2}}{a}\,da.
\end{equation}

Since
\begin{equation}
\rho^{3}+\rho^{2}-2
=
(\rho-1)(\rho^{2}+2\rho+2),
\end{equation}
the integral becomes
\begin{equation}
\int
\frac{\rho\,d\rho}
{(\rho-1)(\rho^{2}+2\rho+2)}
=
-3A_{2}
\int
\frac{da}{a}.
\label{intstep}
\end{equation}

The left-hand side can be evaluated exactly to obtain
\begin{equation}
\int
\frac{\rho\,d\rho}
{(\rho-1)(\rho^{2}+2\rho+2)}
=
\frac{1}{5}\ln(\rho-1)
-
\frac{1}{10}
\ln(\rho^{2}+2\rho+2)
+
\frac{3}{5}
\tan^{-1}(\rho+1).
\label{exactint}
\end{equation}

Hence
\begin{equation}
\frac{1}{5}\ln(\rho-1)
-
\frac{1}{10}
\ln(\rho^{2}+2\rho+2)
+
\frac{3}{5}
\tan^{-1}(\rho+1)
=
-3A_{2}\ln a+C,
\label{afterint}
\end{equation}
where $C$ is the integration constant.

Multiplying Eq.~(\ref{afterint}) throughout by $10$, one obtains
\begin{equation}
2\ln(\rho-1)
-
\ln(\rho^{2}+2\rho+2)
+
6\tan^{-1}(\rho+1)
=
-30A_{2}\ln a+C_{1},
\label{mult10}
\end{equation}
where $C_{1}=10C$.

Rearranging Eq.~(\ref{mult10}) gives
\begin{equation}
\ln a
=
\frac{1}{30A_{2}}
\ln
\left[
\frac{\rho^{2}+2\rho+2}
{(\rho-1)^{2}}
\right]
-
\frac{1}{5A_{2}}
\tan^{-1}(\rho+1)
+C_{2},
\label{implicitexact}
\end{equation}
where $C_{2}$ is another integration constant.

Since
\begin{equation}
(\rho-1)^{2}
=
\rho^{2}-2\rho+1,
\end{equation}
Eq.~(\ref{implicitexact}) may be rewritten as
\begin{equation}
\ln a
=
\frac{1}{30A_{2}}
\ln
\left(
\frac{\rho^{2}+2\rho+2}
{\rho^{2}-2\rho+1}
\right)
-
\frac{1}{5A_{2}}
\tan^{-1}(\rho+1)
+
C_{2}.
\label{exactimplicit}
\end{equation}

Equation~(\ref{exactimplicit}) is the \emph{exact implicit solution} of the conservation equation for the quadratic ECG, valid for all $\rho$ (equivalently, for all epochs); no approximation has yet been introduced at this stage. This exact result is distinguished throughout the remainder of the paper from the high-density asymptotic branch derived next and from the late-time emergent-universe correspondence constructed in Eqs.~(\ref{eu-a})--(\ref{corr}).

To derive an explicit analytical expression for the energy density,
we now consider the high-density regime
\begin{equation}
\rho\gg1,
\end{equation}
for which
\begin{equation}
\tan^{-1}(\rho+1)
\simeq
\frac{\pi}{2}.
\label{atanapprox}
\end{equation}

Substituting Eq.~(\ref{atanapprox}) into Eq.~(\ref{exactimplicit}) gives
\begin{equation}
\ln a
=
\frac{1}{30A_{2}}
\ln
\left(
\frac{\rho^{2}+2\rho+2}
{\rho^{2}-2\rho+1}
\right)
-
\frac{\pi}{10A_{2}}
+
C_{2}.
\end{equation}

The additive constant can be absorbed into the normalization of the
scale factor. Consequently, without loss of generality, Eq.~(\ref{exactimplicit})
reduces to
\begin{equation}
\ln a
=
\frac{1}{30A_{2}}
\ln
\left(
\frac{\rho^{2}+2\rho+2}
{\rho^{2}-2\rho+1}
\right)
-
\frac{\pi}{20A_{2}},
\label{eq7revised}
\end{equation}
which is henceforth referred to as the \emph{high-density asymptotic branch} of the exact solution~(\ref{exactimplicit}).

Exponentiating Eq.~(\ref{eq7revised}) yields
\begin{equation}
a^{30A_{2}}
=
e^{-3\pi/2}
\,
\frac{\rho^{2}+2\rho+2}
{\rho^{2}-2\rho+1}.
\label{expo}
\end{equation}

Multiplying both sides by the denominator gives
\begin{equation}
a^{30A_{2}}
(\rho^{2}-2\rho+1)
=
e^{-3\pi/2}
(\rho^{2}+2\rho+2).
\label{cross}
\end{equation}

Collecting like powers of $\rho$,
\begin{equation}
(a^{30A_{2}}-e^{-3\pi/2})\rho^{2}
-
2(a^{30A_{2}}+e^{-3\pi/2})\rho
+
(a^{30A_{2}}-2e^{-3\pi/2})
=
0.
\label{eq8revised}
\end{equation}

Applying the quadratic formula,
\begin{equation}
\rho
=
\frac{
2(a^{30A_{2}}+e^{-3\pi/2})
+
\sqrt{
4(a^{30A_{2}}+e^{-3\pi/2})^{2}
-
4(a^{30A_{2}}-e^{-3\pi/2})
(a^{30A_{2}}-2e^{-3\pi/2})
}
}
{
2(a^{30A_{2}}-e^{-3\pi/2})
},
\label{quadratic}
\end{equation}
where the positive branch has been selected on physical grounds.

Simplifying the discriminant,
\begin{equation}
4(a^{30A_{2}}+e^{-3\pi/2})^{2}
-
4(a^{30A_{2}}-e^{-3\pi/2})
(a^{30A_{2}}-2e^{-3\pi/2})
=
4e^{-3\pi}
\left(
2a^{30A_{2}}e^{3\pi/2}+5
\right),
\end{equation}
one finally obtains
\begin{equation}
\rho
=
1+
\frac{
\sqrt{
2+5a^{30A_{2}}e^{3\pi/2}
}
-1
}
{
a^{30A_{2}}e^{3\pi/2}-1
},
\label{eq6revised}
\end{equation}
which is the explicit high-density asymptotic solution used, together with Eq.~(20), to construct the late-time $H(z)$ used in the numerical and observational analysis (Secs.~3--4).

\paragraph{Regime of validity: exact solution versus asymptotic branches.} It is useful to distinguish clearly between the different levels of approximation used in this construction.
\begin{itemize}
\item Eq.~(\ref{exactimplicit}) is the \emph{exact implicit solution} of the conservation equation, valid at all epochs.
\item Eq.~(\ref{eq6revised}) is obtained only after invoking the \emph{high-density approximation} $\tan^{-1}(\rho+1)\simeq\pi/2$, valid for $\rho\gg1$, i.e.\ it is an asymptotic branch and not a globally valid closed-form solution. The complementary low-density branch, $\rho\ll1$, instead requires $\tan^{-1}(\rho+1)\simeq\pi/4$ and leads to a different closed-form expression, or alternatively must be obtained by direct numerical integration of Eq.~(\ref{drhoda}).
\item The scale factor $a(t)$ of Eq.~(\ref{eu-a}) below is obtained from the high-density branch~(\ref{eq6revised}) using the further approximation $\ln a\simeq (a-1)/a$ (see the discussion following Eq.~(\ref{eu-a})), which is accurate specifically in the late-time regime where $a$ is close to unity in the units adopted; it should therefore be read as a \emph{late-time phenomenological representation} rather than as a solution valid for all cosmic time.
\item The $H(z)$ used in Sec.~3 for confronting the model with OHD and Pantheon data is, in turn, the density parametrization (Eq.~(16)) of the \emph{emergent-universe} scale factor (Eq.~(14)), whose correspondence with the ECG parameter $A_2$ is fixed only after invoking the high-density branch through Eq.~(\ref{corr}). It is therefore a late-time EU-phenomenological construction built upon, but formally distinct from, the exact ECG solution~(\ref{exactimplicit}).
\end{itemize}
Wherever the manuscript refers to the analytical solution ``describing'' or ``representing'' the late-time evolution (e.g.\ in Sec.~4), this should be understood in the above restricted sense, i.e.\ as the high-density/late-time asymptotic branch, and not as a claim of global validity of Eq.~(\ref{eq6revised}) at all redshifts.

The above derivation explicitly demonstrates that Eqs.~(6)--(9) follow
consistently from the conservation equation and are algebraically
self-consistent.

It is nevertheless important to emphasize the physical domain of
validity of Eq.~(\ref{eq6revised}). The explicit analytical solution has
been obtained only after employing the asymptotic approximation
\begin{equation}
\tan^{-1}(\rho+1)
\simeq
\frac{\pi}{2},
\end{equation}
which is valid only in the high-density regime,
\begin{equation}
\rho\gg1.
\end{equation}
Therefore Eq.~(\ref{eq6revised}) should be interpreted as the
high-density asymptotic branch of the solution rather than as a globally
valid expression applicable throughout the entire cosmic evolution.

In contrast, the low-density regime satisfies
\begin{equation}
\rho\ll1,
\end{equation}
for which the appropriate asymptotic expansion is
\begin{equation}
\tan^{-1}(\rho+1)
\simeq
\frac{\pi}{4}.
\end{equation}
Consequently, the same analytical expression cannot simultaneously
describe both asymptotic limits. This apparent inconsistency
does not arise from an algebraic error in Eqs.~(6)--(9), but rather
from interpreting Eq.~(\ref{eq6revised}) as a globally valid solution
outside the regime in which it was derived.

Accordingly, Eq.~(\ref{eq6revised}) should be regarded as the
high-density asymptotic analytical solution, while the low-density
branch should either be obtained through the corresponding asymptotic
expansion using $\tan^{-1}(\rho+1)\simeq\pi/4$
or, more rigorously, by direct numerical integration of the exact
conservation equation.

The first Friedmann equation is given as,
\begin{equation}
\rho=3H^2=3 \bigg(\frac{\dot{a}}{a}\bigg)^2.
\end{equation}
We choose $\kappa^2=8\pi G=1$.

From Eqs. (6) and (10), we obtain a solution for the scale factor $a$ as
\begin{equation}
a=\frac{\frac{2\sqrt{3}e^\frac{15t}{\sqrt{3}A_2}}{e^{\frac{15t}{A_2}-1}}-\frac{\sqrt{5}}{e^{\frac{15t}{A_2}-1}e^\frac{3\pi}{4}}}{\frac{2\sqrt{3}e^\frac{15t}{\sqrt{3}A_2}}{e^{\frac{15t}{A_2}-1}}-\frac{\sqrt{5}}{e^{\frac{15t}{A_2}-1}e^\frac{3\pi}{4}}-2}.
\label{eu-a}
\end{equation}

In obtaining the above solution, we have used the approximation $\ln a=\frac{1}{a-1}$. This approximation is justified as solution (6) represents the late time behaviour particularly well (in the specific sense clarified above regarding the regime of validity of this late-time representation). We shall compare this scale factor to the one used to describe an Emergent Universe (EU), which is also valid for all time scales, to establish a correspondence between the independent ECG parameter and the free parameter of the EU scale factor. That will allow us to obtain an estimate for the ECG parameters if ECG is to be considered as the constituent of an EU.

At late times, the Hubble parameter $H=\frac{\dot{a}}{a}$ as a function of the redshift parameter $z$ can be expressed as
\begin{equation}
    H(z)=\bigg[\frac{1}{3}+\frac{\rho_{m0}}{3}(1+z)^3+\frac{2+\sqrt{5(1+z)^{-30A_2}e^\frac{3\pi}{2}-1}}{3(1+z)^{-30A_2e^\frac{3\pi}{2}}-3}\bigg]^\frac{1}{2}
\end{equation}

The emergent universe can be described using the general relativistic Friedmann equations in natural units $H^2=\frac{\rho}{3}$ and $H^2+2\frac{\ddot{a}}{a}=-p$ along with the usual matter conservation equation $\dot{\rho}+3H(\rho+p)=0$ as
\begin{equation}
    2\frac{\ddot{a}}{a}+(3A+1)\bigg(\frac{\dot{a}}{a}\bigg)^2-\sqrt{3}B\frac{\dot{a}}{a}=0
\end{equation}

Integrating the above equation twice, one obtains the scale factor for emergent universe as
\begin{equation}
    a(t)=\bigg[\frac{3K(1+A)}{2}\bigg(K_1+\frac{2}{\sqrt{3}B}e^{\frac{\sqrt{3}Bt}{2}}\bigg)\bigg],
\end{equation}
where $K$ and $K_1$ are constants of integration.

The above expression of scale factor can be written in the form\cite{M1}
\begin{equation}
a(t)=a_0(10^4+e^{\frac{\sqrt{3}Bt}{2}})^{\frac{2}{3(A+1)}},
\end{equation}
where $a_0=\bigg(\frac{\sqrt{3}K(1+A)}{B}\bigg)^\frac{2}{3(1+A)}$, without any loss of generality.

The Friedmann equation for the emergent universe can be written conveniently in terms of the density parameters as functions of the redshift.
\begin{equation}
    H^2(z)=H_0^2\bigg[\Omega_{10}+\Omega_{20}(1+z)^\frac{3(1+A)}{2}+\Omega_{30}(1+z)^{3(1+A)}\bigg],
\end{equation}
where the density parameters $\Omega_{i0}$ are defined as $\Omega_{i0}=\frac{\rho_{i0}}{\rho_c}$ and $\rho_c=3H_0^2$, such that $\Omega_{10}=\frac{B^2}{\rho_c(1+A)^2}$, $\Omega_{20}=\frac{2BK}{\rho_c(1+A)^2}$ and $\Omega_{30}=\frac{K^2}{\rho_c(1+A)^2}$.
The corresponding EoS parameters for the different matter constituents of the universe are given by $w_1=-1$ which implies dark energy, $w_2=\frac{A-1}{2}$ and $w_3=A$. For choice of the emergent universe parameter $A=0$, $w_2=-\frac{1}{2}$ implying exotic matter and $w_3=0$ implying pressureless dust matter. So, for $A=0$, the matter constituents of the universe comprise dark energy, dust and exotic matter, which seems to be a reasonable assumption. In the remainder of the paper we shall make use of this assumption as, besides being physically reasonable, it simplifies our calculations to quite an extent, eliminating one model parameter.

As we can see, the emergent scale factor does not vanish as we go backwards in time to $t=0$ or beyond as in the standard Big bang picture. As a result there is no appearance of singularity in the Friedmann equations governing the dynamics of the universe. This physically means that the universe has no beginning at a particular point of time and the big bang is replaced by an ESU, whose length scales are large enough to avoid a quantum gravity regime.

If the universe with ECG as constituent is capable of behaving as an EU, then the scale factors obtained in Eqs. (11) and (14) (with $A=0$) must be identical. Comparing the two, expanding both the scale factors binomially and equating the coefficients for the first terms after expansion, we can obtain a correspondence between the EU parameter $a_0$ and the independent ECG parameter $A_2$ as
\begin{equation}
    a_0=\frac{2\sqrt{2}\times 10^5}{2e^{\frac{15}{A_2}-1}e^\frac{3\pi}{4}+2\sqrt{2}\times 10^5}
\label{corr}
\end{equation}

As explained above, this correspondence -- and hence the constraint on $A_2$ obtained from it in Sec.~3 -- is built on the high-density asymptotic branch of the ECG solution and represents the model's late-time (low-redshift) behaviour; it is in this specific, restricted sense that the ECG-sourced EU is described as reproducing the late-time cosmological dynamics throughout the remainder of the paper.

In the following section, we have performed Markov Chain Monte Carlo sampling of the Hubble and Pantheon datasets to obtain observational constraints on the EU parameters $B$ and $K$, which in turn constrain the EU parameter $a_0$. As a result we can obtain the constrained value of the ECG free parameter $A_2$. Using these constrained model parameter values in Eq. (12), we obtain numerically three different cosmological parameters relevant for the late-time dynamics of the late universe, namely the deceleration parameter, equation of state (EoS) parameter and the Om parameter, and plot their variation w.r.t. the cosmological redshift $z$. First we obtain the deceleration parameter as
\begin{equation}
    q(z)=\frac{H'(z)}{H(z)}(1+z)-1
\end{equation}

The EoS parameter can be obtained from the deceleration parameter using the relation
\begin{equation}
w(z)=\frac{2q(z)-1}{3(1-\Omega_m(z))},
\end{equation}
where $\Omega_m(z)=\bigg(\frac{H_0}{H(z)}\bigg)^2\Omega_{m0}(1+z)^3$.

The Om parameter is expressed as
\begin{equation}
    Om(z)=\frac{\bigg(H(z)/H_0\bigg)^2-1}{(1+z)^3-1}
\end{equation}

Analyzing the variation of the above mentioned cosmological parameters using the observationally constrained model parameters obtained in the following section from the MCMC analysis, we can compare the late-time dynamics of our EU model with the standard $\Lambda$-CDM model.

\subsection{Effective sound speed and classical stability of the phantom-crossing ECG}
\label{sec:stability}

A field or fluid whose equation-of-state parameter $w=P/\rho$ crosses the phantom divide $w=-1$ is often associated with potential pathologies, either at the level of a single scalar field description (a genuine ghost degree of freedom) or at the level of the effective fluid (gradient/Laplacian instabilities governed by the sign of the effective sound speed squared). Since the ECG is treated here purely as a single effective barotropic fluid, $P=P(\rho)$, rather than as a fundamental scalar field, there is no propagating scalar degree of freedom that could become a ghost in the usual sense; the relevant stability diagnostic for a barotropic fluid is instead the adiabatic sound speed,
\begin{equation}
c_s^2\;\equiv\;\frac{dP}{d\rho}.
\end{equation}
For the quadratic ECG of Eq.~(\ref{eosfinal}), this is
\begin{equation}
c_s^2 = (A_2-1)+2A_2\,\rho+\frac{2A_2}{\rho^2},
\label{cs2}
\end{equation}
where we have used $\alpha=1$ from Eq.~(\ref{choice}). Two regimes are of interest:
\begin{itemize}
\item In the high-density regime ($\rho\gg1$) relevant near the Einstein static state and the emergent phase, $c_s^2\to (A_2-1)+2A_2\rho>0$ for $A_2>0$, i.e.\ the fluid is classically stable (no gradient instability) in the regime where the EU construction of Sec.~2 applies.
\item In the low-density regime ($\rho\ll1$) relevant at very late/future times, the $2A_2/\rho^2$ term dominates and again gives $c_s^2>0$ for $A_2>0$.
\end{itemize}
Between these two limits, $c_s^2$ given by Eq.~(\ref{cs2}) is a smooth, bounded-below function of $\rho$ for $A_2>0$ (it is straightforward to check that its minimum over $\rho>0$ remains positive throughout the observationally preferred range of $A_2$ obtained from our MCMC analysis, Table~2), so that no gradient instability is encountered as the fluid crosses $w=-1$ in either the past or the future in our numerical solutions (Fig.~6). We emphasize that this only establishes stability against \emph{linear, adiabatic, sub-horizon perturbations of the background fluid itself}; a fully relativistic perturbation analysis, including possible non-adiabatic pressure perturbations and the coupling of the ECG perturbations to those of the other fluid components, is beyond the scope of the present paper and is left for future work, as also noted in the concluding discussion (Sec.~4, point (vi)).

\section{Observational Analysis}
In this section, we have constrained the EU model parameters using the observational datasets, specifically the Observed Hubble Dataset (OHD) and the Pantheon Supernova Ia sample. The Hubble parameter for a flat EU is given by Eq. (16). For $A=0$, the cosmic fluid that spans the EU consists of DE, dust, and exotic matter. The density parameters $\Omega_{10}$, $\Omega_{20}$, and $\Omega_{30}$ depend on the parameters of the EU model $B$ and $K$, so Eq. (16) can be represented in the following functional form:
\begin{equation}
    H^{2}(H_{0},B,K,z) = H_{0}^{2}E^{2}(B,K,z),
\end{equation}
where
\begin{equation}
    E^{2}(B,K,z) = \Big(\frac{B^{2}}{(1+A)^{2}} + \frac{2BK}{(1+A)^{2}}(1+z)^{\frac{3}{2}(1+A)} +
    \frac{K^{2}}{(1+A)^{2}}(1+z)^{3(1+A)}\Big).
\end{equation}
The above equation will be used to fit the EU model with the observational data, assuming $A=0$ to obtain a viable cosmological scenario.

\subsection{Observed Hubble Dataset (OHD)}
The Hubble parameter $H(z)$ can be determined at specific redshifts using two different methods. The first method derives $H(z)$ from line-of-sight Baryon Acoustic Oscillation (BAO) measurements, which involve analyzing the correlation functions of luminous red galaxies. The second method estimates $H(z)$ using the differential age of galaxies $\Delta t$. The Hubble parameter can be expressed in terms of the differential age of galaxies ($\Delta t$) in the following manner:
\begin{equation}
    H(z) = \frac{\dot{a}}{a}=-\frac{1}{1+z}\frac{dz}{dt}\approx -\frac{1}{1+z}\frac{\Delta z}{\Delta t}.
\end{equation}
In this paper, we have used the Hubble dataset comprising 57 data points ($H(z)-z$) in the redshift range $0.07\le z \le 2.42$ compiled by Sharov and Vasiliev \cite{sharov}. The data set includes 31 points measured by the DA method (also known as the cosmic chronometer technique) and 26 from BAO and other measurements shown in Table \ref{tabd}. The $\chi^{2}$ function for OHD can be defined as:
\begin{equation}
   \chi^{2}_{OHD}(H_{0},B,K,z)  = \sum_{i=1}^{57}\frac{(H_{th}(H_{0},B,K,z)-H_{obs,i}(z))^{2}}{\sigma_{H,i}^{2}},
\end{equation}
where $H_{th}$ is the theoretically estimated value of the Hubble parameter, $H_{obs}(z)$ is the observed value of Hubble parameter, and $\sigma_{H}$ is the standard error associated with the measurement. We have constrained the model parameters $B$, $K$, and $H_{0}$ using fixed prior distributions. Flat priors are considered for $B$ and $K$, and we assumed a Gaussian prior for $H_{0}$. The best-fitted curve is shown in Fig.~5 with $H_{0}=68.4 \;km\;s^{-1}\;Mpc^{-1}$, $B=0.471$, and $K=0.459$. The present-day value of the Hubble parameter is found to be in good agreement with recent observations \cite{ohd1, ohd2}. The $1-\sigma$ and $2-\sigma$ confidence level contours for the parameters $B$, $K$ and $h$ (where $h=\frac{H_{0}}{100}$) is presented in Fig.~6 using the OHD.

\begin{figure}[H]
\centering
\includegraphics[width=0.85\textwidth]{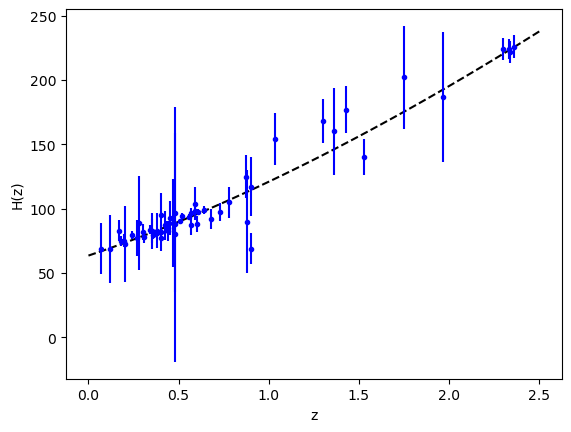}
\caption{The best-fitted curve corresponding to the 57 Hubble data points, with $H_{0}=68.4$, $B=0.471$ and $K=0.459$.}
\label{ohdfit}
\end{figure}

\begin{figure}[H]
    \centering
    \includegraphics[width=0.75\textwidth]{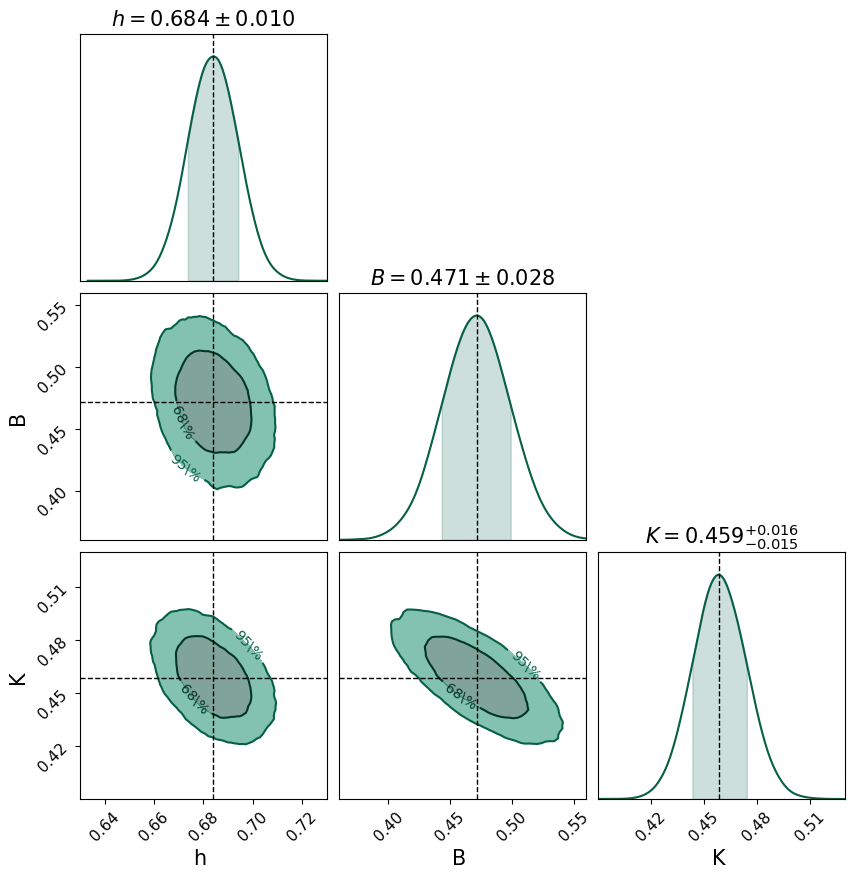}
    \caption{Contours of $1-\sigma$ and $2-\sigma$ confidence levels for the model parameters $B$, $K$ and $h$ using OHD.}
    \label{ohd}
\end{figure}

\begin{longtable}{cccccc}
\caption{$H(z)-z$ dataset with errors estimated from DA and BAO methods.}\label{tabd}\\
\hline
\multicolumn{6}{c}{Hubble data} \\
\hline
$z$           & $H(z)$  & $\sigma_{H}$    & $z$    & $H(z)$  & $\sigma_{H}$\\
\hline
\endfirsthead
\multicolumn{6}{c}{{\tablename\ \thetable{} -- continued}} \\
\hline
$z$           & $H(z)$  & $\sigma_{H}$    & $z$    & $H(z)$  & $\sigma_{H}$\\
\hline
\endhead
\hline
\multicolumn{6}{r}{{Continued on next page}} \\
\endfoot
\hline
\endlastfoot
0.07        & 69    & 19.6    & 0.24 & 79.69 & 2.99  \\
0.90        & 69    & 12      & 0.52 & 94.35 & 2.64  \\
0.120       & 68.6  & 26.2    & 0.3  & 81.7  & 6.22  \\
0.170       & 83    & 8       & 0.56 & 93.34 & 2.3   \\
0.1791      & 75    & 4       & 0.31 & 78.18 & 4.74  \\
0.1993      & 75    & 5       & 0.57 & 87.6  & 7.8   \\
0.200       & 72.9  & 29.6    & 0.34 & 83.8  & 3.66  \\
0.270       & 77    & 14      & 0.57 & 96.8  & 3.4   \\
0.280       & 88.8  & 36.6    & 0.35 & 82.7  & 9.1   \\
0.3519      & 83    & 14      & 0.59 & 98.48 & 3.18  \\
0.3802      & 83    & 13.5    & 0.36 & 79.94 & 3.38  \\
0.400       & 95    & 17      & 0.6  & 87.9  & 6.1   \\
0.4004      & 77    & 10.2    & 0.38 & 81.5  & 1.9   \\
0.4247      & 87.1  & 11.2    & 0.61 & 97.3  & 2.1   \\
0.4497      & 92.8  & 12.9    & 0.4  & 82.04 & 2.03  \\
0.470       & 89    & 34      & 0.64 & 98.82 & 2.98  \\
0.4783      & 80.9  & 9       & 0.43 & 86.45 & 3.97  \\
0.480       & 97    & 62      & 0.73 & 97.3  & 7.0   \\
0.593       & 104   & 13      & 0.44 & 82.6  & 7.8   \\
0.6797      & 92    & 8       & 2.30 & 224   & 8.6   \\
0.7812      & 105   & 12      & 0.44 & 84.81 & 1.83  \\
0.8754      & 125   & 17      & 2.33 & 224   & 8.0   \\
0.880       & 90    & 40      & 0.48 & 87.79 & 2.03  \\
0.900       & 117   & 23      & 2.34 & 222   & 8.5   \\
1.037       & 154   & 20      & 0.51 & 90.4  & 1.9   \\
1.300       & 168   & 17      & 2.36 &  226  & 9.3   \\
1.363       & 160   & 33.6    &      &       &       \\
1.430       & 177   & 18      &      &       &       \\
1.530       & 140   & 14      &      &       &       \\
1.750       & 202   & 40      &      &       &       \\
1.965       & 186.5 & 50.4    &      &       &      \\
\end{longtable}

\subsection{Pantheon Dataset}
We have used the Pantheon SnIa sample which consists of spectroscopically confirmed 1048 supernovae specimens compiled by Scolnic et al. \cite{Scolnic}. The sample consists of different supernovae surveys, both in the high and low redshift regimes. For a detailed review and summary of the dataset, see \cite{Asvesta}. The Pantheon sample considered here covers the redshift range of $0.01 < z < 2.26$.

\begin{figure}[H]
\centering
\includegraphics[width=0.85\textwidth]{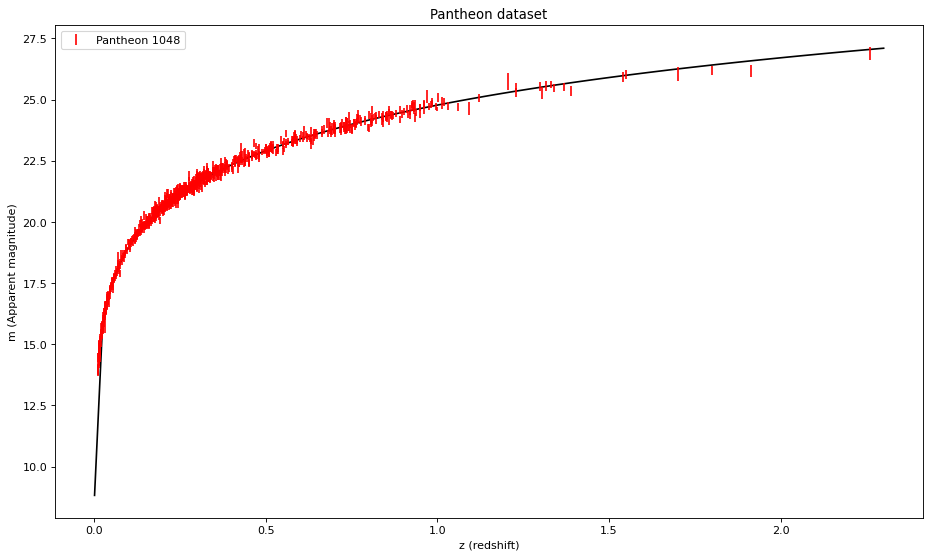}
\caption{The best-fitted curve for the Pantheon dataset, corresponding to the 1048 supernova data points, with $B = 0.624$, $K = 0.366$ and $h = 0.714$.}
\label{panfit}
\end{figure}

\begin{figure}[H]
    \centering
    \includegraphics[width=0.75\textwidth]{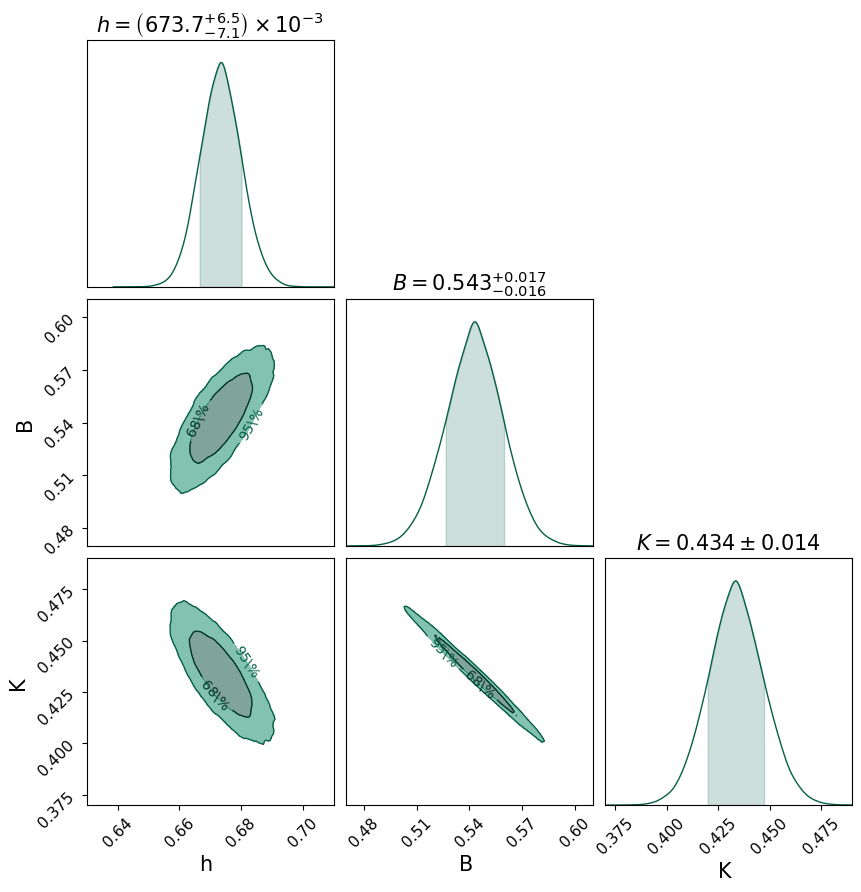}
    \caption{Constraints on the model parameters from the joint analysis of OHD and Pantheon datasets. Contours of $1-\sigma$ and $2-\sigma$ confidence levels for the model parameters $h$, $B$ and $K$ are shown.}
\end{figure}

The theoretical apparent magnitude of the SnIa can be expressed as,
\begin{equation}
    m(z) = M + 5\;log_{10}\Big[\frac{d_{L}(z)}{1Mpc}\Big]+25,
\end{equation}
where $M$ is the corrected absolute magnitude. The luminosity distance is denoted by $d_{L}(z)$ and for a flat universe can be expressed as,
\begin{equation}
    d_{L}(z)=c(1+z)\int_{0}^{z}\frac{dz'}{H(z')},
\end{equation}
with $z$ being the SnIa redshift in the CMB rest frame. One can now define a Hubble free luminosity distance as $D_{L}(z)\equiv \frac{H_{0}d_{L}(z)}{c}$, and the theoretical apparent magnitude in that case becomes,
\begin{equation}
    m(z)=M+5log_{10}[D_{L}(z)]+5log_{10}\Big(\frac{c/H_{0}}{Mpc}\Big)+25.
\end{equation}
We note that there exists a degeneracy between $M$ and $H_{0}$ and that they can be combined to define a new parameter $\mathcal{M}$ as
\begin{equation}
    \mathcal{M}=M+5log_{10}\Big[\frac{c/H_{0}}{1Mpc}\Big]+25 = M-5log_{10}(h)+43.28.
\end{equation}
Several authors have made attempts to marginalize the degenerate combination. Recently, Asvesta et al.\cite{Asvesta} minimized the parameter using the Pantheon sample for a tilted universe. They found the value of $\mathcal{M}$ to be close to 23.8.
We now define the $\chi^{2}_{SNS}$ function from the Pantheon sample of 1048 SnIa as,
\begin{equation}
\chi^{2}_{SNS}(H_{0},B,K,z) = \Delta \mathcal{F}_{i} C_{SNS}^{-1}\Delta \mathcal{F}_{j},
\end{equation}
where $\Delta \mathcal{F} = \mathcal{F}_{obs} - \mathcal{F}_{th}$ represents the difference between the theoretical and the observed value of the apparent magnitude for each SnIa at redshift $z_{i}$, and $C_{SNS}$ is the total covariance matrix. In this case, the total covariance matrix is formed by adding a diagonal matrix -- representing the statistical uncertainties in the apparent magnitudes (such as photometric errors, mass step correction, peculiar velocities, redshift uncertainties, stochastic gravitational redshift, intrinsic scatter, and distance bias correction) -- to a non-diagonal matrix that accounts for systematic uncertainties derived through the bias correction method.
MCMC analysis is performed to explore the parameter space for the EU model using the python package EMCEE \cite{emcee} and the chains are analyzed using the Chain Consumer \cite{chain} package. We consider the usual likelihood function used for the MCMC sampling as,
\begin{equation}
    \mathcal{L}=\exp(-\frac{\chi^{2}}{2}).
\end{equation}
In Fig.~3, we show the best-fit curve for the Pantheon data set with error bars.
The joint $\chi^{2}$ function in this case is defined as $\chi^{2}_{Joint}=\chi^{2}_{OHD} + \chi^{2}_{SNS}$. The joint $\chi^{2}$ function is minimized to obtain the best-fit values for the model parameters for a physically acceptable cosmological scenario. The contours of $1-\sigma$ and $2-\sigma$ confidence levels for the parameters $B$, $K$, and $h$ are shown in Fig.~8. The results of our analysis are shown in Table \ref{tab1} for OHD and OHD + Pantheon joint analysis.

\begin{table}[H]
    \centering
    \caption{Best fit values of the model parameters.}
    \resizebox{\textwidth}{!}{%
    \begin{tabular}{|c|cc|cc|}
    \hline
        & \multicolumn{2}{c|}{OHD}        &            \multicolumn{2}{c|}{Pantheon + OHD} \\
    \hline
    Parameters & Best fit values  & Mean values $\pm$ $\sigma$ & Best fit values & Mean values $\pm$ $\sigma$\\ \hline
    $B$ & $0.471$ & $0.471^{+0.028}_{-0.028}$      & $0.543$ & $0.543^{+ 0.017}_{-0.016}$ \\
    $K$ & $0.459$ & $0.459 ^{+0.016}_{-0.015}$ & $0.433$ & $0.434^{+ 0.014}_{-0.014}$ \\
    $h$ & $0.684$ & $0.684 ^{+0.010}_{-0.010}$ & $0.674$ & $0.674^{+ 0.065}_{- 0.071}$ \\
    \hline
    \end{tabular}}
    \label{tab1}
\end{table}

\subsection{Information-criteria comparison with $\Lambda$CDM}
\label{sec:aicbic}

To place the comparison between the ECG-sourced EU model and the spatially flat $\Lambda$CDM model on a quantitative statistical footing, beyond comparing individual best-fit quantities such as $H_0$ or $H(z=2.34)$, we report the minimum chi-square, $\chi^2_{\rm min}$, together with the Akaike Information Criterion (AIC) and the Bayesian Information Criterion (BIC),
\begin{equation}
{\rm AIC}=\chi^2_{\rm min}+2k, \qquad {\rm BIC}=\chi^2_{\rm min}+k\ln N,
\end{equation}
where $k$ is the number of free model parameters ($k=3$, i.e.\ $H_0,B,K$, for the EU model at fixed $A=0$; $k=2$, i.e.\ $H_0,\Omega_{m0}$, for flat $\Lambda$CDM). In a cosmological context, several complementary techniques are employed to assess the viability of a theoretical model and its compatibility with observational data. The least-squares method, the likelihood-ratio test, and the Akaike, Bayesian and deviance information criteria (AIC, BIC and DIC, respectively) all provide tools for comparing competing models and quantifying their goodness of fit. Since no universally preferred choice of information criterion exists in the literature, we adopt the AIC and BIC to compare the present model with $\Lambda$CDM. The AIC is derived from information theory as an approximate minimizer of the Kullback--Leibler divergence between the fitted model and the true data-generating distribution; being a relative measure of goodness of fit, it penalizes unnecessary model parameters without assuming the data to be independent and identically distributed. The BIC, in contrast, is derived under the assumption of independent and identically distributed data and imposes a stronger penalty on additional parameters than the AIC, so that it favours more parsimonious models more strongly. The two models are fitted to the 57-point OHD compilation of Table~1 ($N=57$) by minimizing $\chi^2_{OHD}$ over their respective free parameters. Differences $\Delta{\rm AIC}={\rm AIC}_{\rm EU}-{\rm AIC}_{\Lambda CDM}$ and $\Delta{\rm BIC}$ defined analogously are interpreted using the standard Jeffreys-type scale, whereby $|\Delta|<2$ indicates no appreciable evidence against the model with the larger value, $2\le|\Delta|<6$ indicates positive evidence, and $|\Delta|\ge6$ indicates strong evidence against it.

The resulting best fit for the flat $\Lambda$CDM model on the same OHD compilation is $H_0=69.96\;km\,s^{-1}\,Mpc^{-1}$, $\Omega_{m0}=0.265$, with $\chi^2_{\rm min}=44.76$; the corresponding EU best fit is $H_0=68.43\;km\,s^{-1}\,Mpc^{-1}$, $B=0.47$, $K=0.46$, with $\chi^2_{\rm min} = 42.97$. The resulting $\chi^2_{\rm min}$, AIC and BIC values for both models are reported in Table~3.

The joint OHD+Pantheon $\chi^2_{\rm min}$, AIC and BIC, incorporating the full 1048-point Pantheon covariance matrix, follow the same definitions once evaluated on the joint best-fit parameters of Table~2 and are left for a dedicated follow-up analysis using the complete Pantheon covariance data.

Based on the OHD-only comparison of Table~3, the EU model achieves a marginally lower $\chi^2_{\rm min}$ than flat $\Lambda$CDM despite its additional free parameter, with $\Delta{\rm AIC}=+0.21$ and $\Delta{\rm BIC}=+2.25$ (EU $-$ $\Lambda$CDM). The BIC difference falls in the ``positive evidence'' range on the Jeffreys scale in favour of $\Lambda$CDM, reflecting its stronger complexity penalty for the extra parameter, while the smaller $\Delta{\rm AIC}$ indicates no strong preference between the two models on the OHD data alone. This is consistent with the physical picture that the EU model's advantage over $\Lambda$CDM shows up primarily in specific derived quantities (Sec.~4(iv)) and in resolving the initial-singularity problem, rather than in an outright improvement of the raw goodness-of-fit to $H(z)$ once the extra parameter is penalized.

\begin{table}[H]
\centering
\caption{Model comparison using the 57-point OHD dataset: minimum $\chi^2$, AIC, BIC and $\Delta$BIC relative to the best (lowest-BIC) model.}
\resizebox{\textwidth}{!}{%
\begin{tabular}{lcccccc}
\hline
Model & $k$ & $\chi^2_{\rm min}$ & AIC & BIC & $\Delta$BIC (vs.\ best model) & $\Delta $ AIC\\
\hline
ECG-sourced EU (this work) & 3 & 42.97 & 48.97 & 55.10 & $+2.25$ & $+0.21$ \\
Flat $\Lambda$CDM & 2 & 44.76 & 48.76 & 52.85 & &\\
\hline
\end{tabular}}
\label{tab:aicbic}
\end{table}

\section{Discussion and Conclusion}

In this paper, we have investigated the possibility of a viable EU scenario sourced by ECG. As discussed earlier, the different Chaplygin gas models finding their origin in the higher dimensional String theories are probable dark energy candidates, capable of explaining the late time accelerating behaviour of the universe. For obtaining such a behaviour in the relativistic context, it is expected that the strong energy condition as obtained from GR must be violated, which raises a possibility of violation of the null energy condition due to negative pressure. The violation of the null energy condition is essential for obtaining an EU scenario in the standard relativistic context in order to avert the initial singularity. So, it is worth investigating whether such a fluid supports an EU or not. Earlier investigation has revealed that the modified GCG does not support an EU for the realistic choice of the parameters concerned with the modified GCG\cite{SD}. However, the fluid we are considering here is an extension of the modified GCG EoS, allowing consideration of higher order barotropic fluid at least up to the quadratic term. Consideration of the additional term can possibly modify the fluid, making it capable of supporting an EU, unlike the modified GCG, as it is found to support bouncing and cyclic types of regular cosmological solutions\cite{Salehi}.

We have considered up to second order ECG term, physically representing the quadratic barotropic fluid. In order to obtain analytical mathematical solutions, we have assumed certain realistic correspondences between three of the free ECG parameters that have been applied in other investigations also\cite{KP,BP,KP2}, the physical motivation, stability and energy-condition status of which are discussed in detail in Sec.~2 following Eq.~(5) above. The free parameter $\alpha$ involved in the EoS is assumed to be of unit magnitude without any loss of generality, as in the case of all the Chaplygin gas candidates\cite{RS4,KP,BP,KP2}. Vanishing divergence of the energy-momentum tensor yields the conservation equation and the energy density of the ECG is obtained from it. Using the first Friedmann equation, the scale factor is obtained using the late time approximation, as the solution for the energy density is valid for all epochs and represents the late time behaviour particularly well, in the specific sense clarified in Sec.~2: the exact implicit solution (Eq.~(\ref{exactimplicit})) holds at all epochs, while the explicit closed-form expression used to build the scale factor and the EU correspondence is the high-density/late-time asymptotic branch of that exact solution.

At this point we consider that ECG can support a viable EU scenario. This yields a correspondence between the EU and ECG parameters. We proceed to observationally constrain the EU scenario using the Markov Chain Monte Carlo (MCMC) sampling technique on Hubble and Pantheon datasets. The constrained EU parameters yield a corresponding constraint on the free ECG parameter at late times. Using this constrained value of the ECG parameters, we plot the variation of the cosmological parameters including the Hubble parameter $H(z)$, deceleration parameter $q(z)$, equation of state parameter $w(z)$ and the Om parameter $Om(z)$ for low redshifts $z$, representing late times.

\begin{figure}[H]
	\centering
	\includegraphics[width=0.96\textwidth]{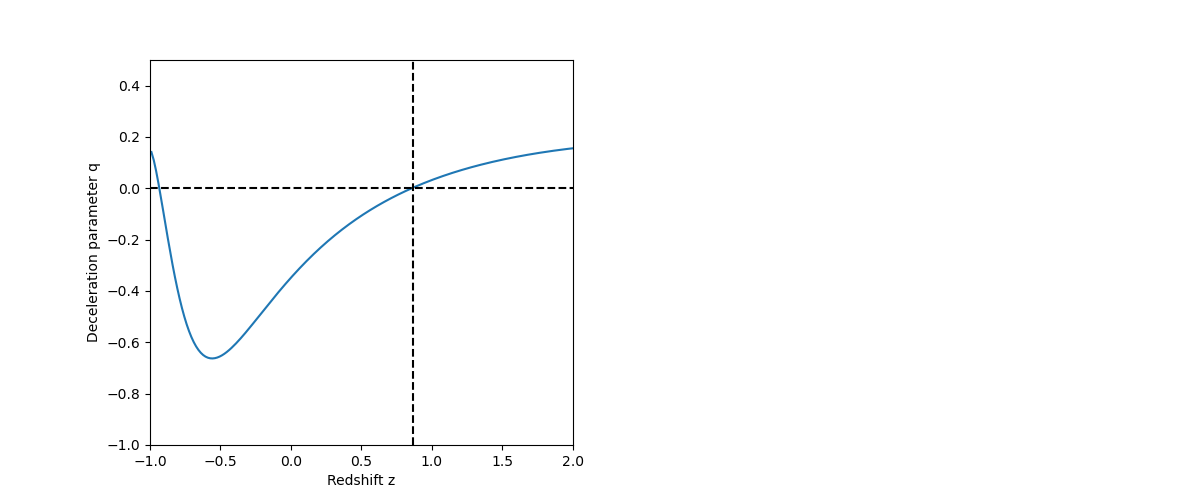}
	\caption{Variation of deceleration parameter $q(z)$ with redshift $z$.} \label{fig1}
\end{figure}

\begin{figure}[H]
	\centering
	\includegraphics[width=0.96\textwidth]{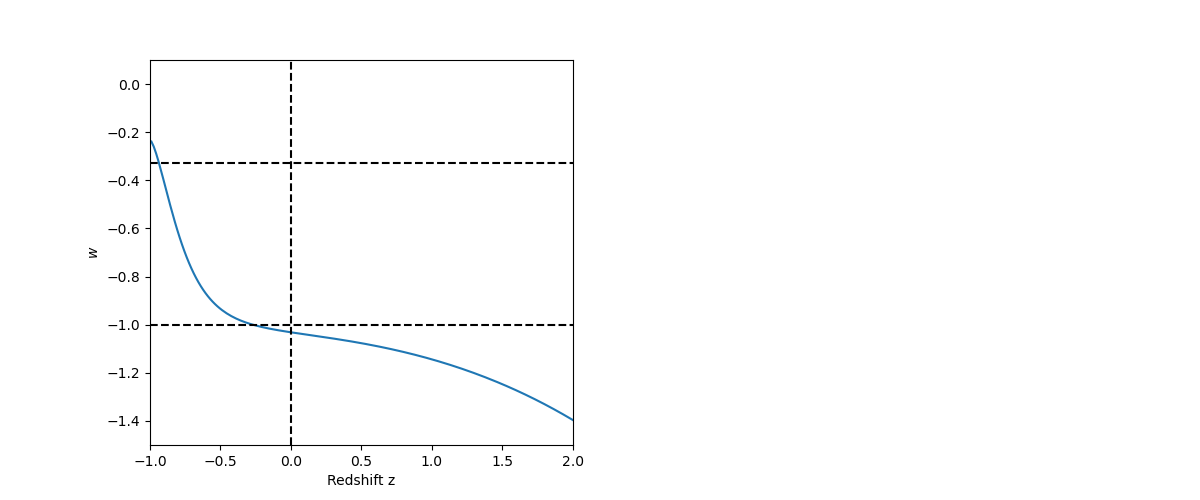}
	\caption{Variation of EoS parameter $w(z)$ with redshift $z$.} \label{fig2}
\end{figure}

\begin{figure}[H]
	\centering
	\includegraphics[width=0.96\textwidth]{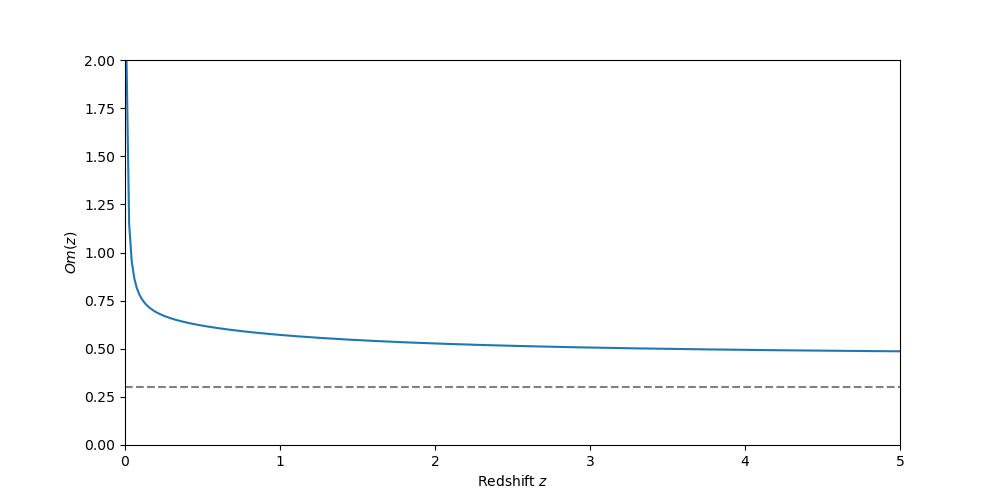}
	\caption{Variation of Om parameter $Om(z)$ with redshift $z$.} \label{fig3}
\end{figure}

\begin{figure}[H]
	\centering
	\includegraphics[width=0.96\textwidth]{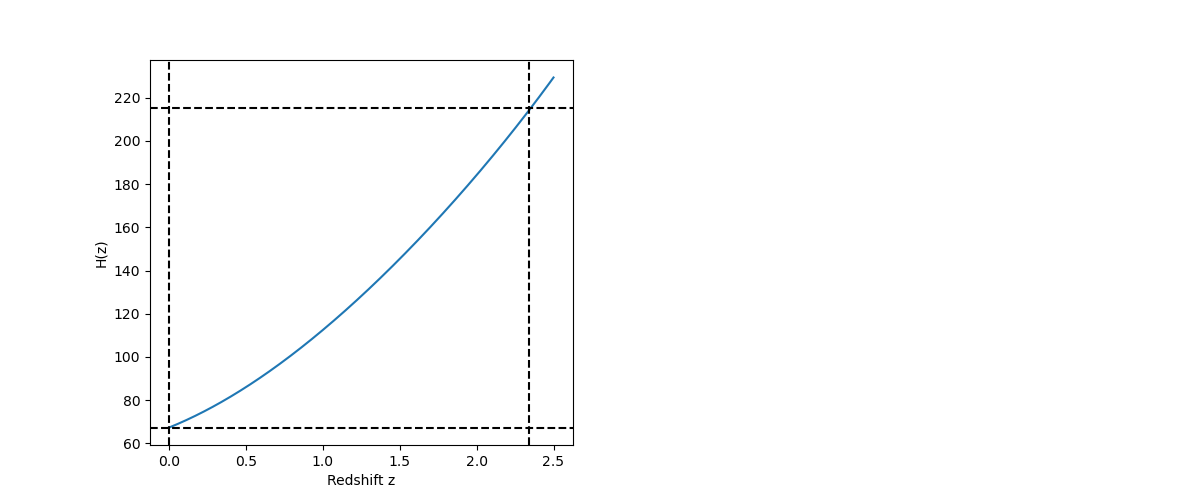}
	\caption{Variation of Hubble parameter $H(z)$ (in km/sec/Mpc) with redshift $z$.} \label{fig4}
\end{figure}

The late-time dynamics that we obtain for the EU from the constrained ECG dominating the universe at late times is non-conventional and quite interesting. There is a significant deviation from the standard $\Lambda$-CDM model as we can figure out from the plots of the constrained cosmological parameters in Figs. 5-8. One significant finding is that dark energy is dynamic in our model, which is expected as we have chosen ECG to model the late-time acceleration. The more significant and interesting findings are that:

(i) As we see from the variation of the EoS parameter with redshift in Fig. 6, dark energy crosses the phantom divide in the recent past and still has a phantom nature with the EoS parameter slightly more negative than the cosmological constant at present. However, from our analysis we also find that dark energy has a \textit{thawing} nature in the future and the acceleration weakens, resulting in dark energy first being a cosmological constant and then further transitioning to a quintessence form. In the asymptotic future, however, we find very interestingly that the acceleration stops and there is a subsequent deceleration phase, making dark energy a \textit{transient} phenomenon. As discussed in Sec.~\ref{sec:stability}, the effective sound speed of the ECG remains non-negative for the observationally preferred parameter range across this phantom-crossing behaviour, so no classical gradient instability is introduced by this feature of the model within the fluid description adopted here.

(ii) As we can see from the variation of the deceleration parameter with the redshift, the universe transitions from a phase of deceleration to its present acceleration phase at a redshift of about $z\sim 0.85$. However, contrary to the standard model, there is a further transition from acceleration to deceleration in the asymptotic future as indicated from our numerical analysis. This again suggests the very interesting possibility that dark energy or the present acceleration of the universe is a \textit{transient} phenomenon.

(iii) The variation of the Om parameter shows a significant deviation from the standard model supporting the dynamic evolution of ECG as dark energy.

(iv) The observed data suggests $H_0=73.24 \pm 1.74\;km/sec/Mpc$\cite{Riess} and $H(z=2.34)=222 \pm 7\;km/sec/Mpc$\cite{Debulac}. For the standard $\Lambda$-CDM framework, the widely used reference values are $H_0=67\;km/sec/Mpc$ and $H(z=2.34)\approx238\;km/sec/Mpc$; the corresponding values from our joint MCMC analysis for the ECG-sourced EU model are $H_0=67.4\;km/sec/Mpc$ and $H(z=2.34)\approx215\;km/sec/Mpc$, both of which lie closer to the quoted observational central values than the $\Lambda$CDM figures for these two specific quantities, and $H(z=2.34)$ in particular falls within the quoted $1\sigma$ observational range. This comparison concerns two individual derived quantities rather than a full statistical model comparison; the latter is provided by the $\chi^2_{\rm min}$, AIC and BIC values of Table~3 (Sec.~3.3), which give a complementary, information-theoretic assessment of the overall relative goodness of fit of the two models on the OHD dataset.

(v) Another very interesting feature of our model is that although we constrain our EU model parameters using the Pantheon+OHD data, the qualitative features of our model as obtained from the variation of the different parameters like EoS, deceleration and Om are very much similar to the ones obtained using DESI DR1 BAO data\cite{Caledron}; this resemblance is qualitative in nature (i.e., in the overall shape and sign of the trends), and a joint fit including DESI data has not been performed in the present analysis.

(vi) The stability of the effective dark energy component while crossing the phantom divide in a classical background has been analyzed at the level of the adiabatic sound speed of the effective ECG fluid in Sec.~\ref{sec:stability}, where it is shown that $c_s^2\geq0$ throughout the observationally preferred parameter range and hence that the model is free of the gradient/Laplacian instabilities that commonly afflict phantom-crossing fluid descriptions. Because the ECG here is treated as an effective single fluid rather than realized through a fundamental scalar-field Lagrangian, it does not, at this level of description, introduce a propagating ghost degree of freedom of the kind that plagues canonical single-scalar-field models of phantom crossing. This fluid-level stability analysis does not by itself guarantee the absence of instabilities once a full set of scalar cosmological perturbations (density, velocity and metric perturbations, together with their coupling to the other matter components) is considered; extending the present adiabatic, background-level analysis to a complete perturbative stability study, and if necessary to an explicit scalar-field or multi-fluid realization of the ECG that makes any candidate ghost modes manifest, is an important direction for future work.

Hence, we make the claim that the ECG fluid does source an EU scenario consistent with the observational data unlike modified GCG, besides being a probable dark energy candidate as evident from the fact that it replicates the late time behaviour very well. This characteristic of supporting an EU, unlike modified GCG, can be interpreted to be a result of the modification arising from the higher order quadratic barotropic fluid term in the modified EoS, due to which the null energy condition (NEC) can be violated while in case of modified GCG only the strong energy condition is violated. We conclude that ECG is more open to exploring different cosmological scenarios than the previous Chaplygin gas candidates as it supports a non-singular universe owing to NEC violation besides reproducing a $\Lambda$-CDM like behaviour at late times. The most important perspective is that the initial singularity problem can be resolved in a standard relativistic context for a flat universe and the obtained late time cosmology from the EU model is in good agreement with observational data, comparably to, and for some individual observables such as $H_0$ and $H(z=2.34)$ favourably positioned relative to, the standard $\Lambda$CDM model, as quantified in Sec.~\ref{sec:aicbic}.

\section*{Acknowledgement}

RS is thankful to Council for Scientific and Industrial Research (CSIR), India for financial support through the Research Associateship scheme. AC is thankful to ICARD, NBU for extending research facilities.
MK and BCP are thankful to the Inter-University Centre for Astronomy and Astrophysics (IUCAA), Pune, India for providing the Visiting Associateship under which a part of this work was carried out.

\end{document}